\documentstyle[12pt]{article}

\def\r{\right}
\def\l{\left}
\def\bq{\begin{equation}}
\def\eq{\end{equation}}

\begin{document}
\begin{flushright}
CINVESTAV-FIS-14/95
\end{flushright}

\vspace{1cm}

\begin{center}
{\bf Update to the Neutrino-Electron Scattering in Left-Right Symmetric Models}
\end{center}

\vspace{1cm}

\begin{center}
O.~G.~Miranda \footnote[1]{e-meil: omr@fis.cinvestav.mx}  \\
Centro de Investigaci\'on y de Estudios Avanzados del I.~P.~N. \\
Dpto.~de F\'{\i}sica, A.~P.~14--740, M\'exico 07000, D.~F., M\'exico.
\end{center}
\begin{center}
M.~Maya \\
Universidad Aut\'onoma de Puebla, \\
Facultad de Ciencias F\'{\i}sico Matem\'aticas
A.~P.~1364, 72000, Puebla,
M\'exico.
\end{center}
\begin{center}
R. Huerta \footnote[2]{e-mail: rhuerta@kin.cieamer.conacyt.mx.} \\
Centro de Investigaci\'on y de Estudios Avanzados del I.~P.~N. \\
U.~Merida, Dpto.~de F\'{\i}sica Aplicada, A.~P.~73,
Cordemex 97310
M\'erida, Yucat\'an, M\'exico.
\end{center}

\vspace{1cm}


\begin{abstract}

In a previous paper we study the neutrino-electron scattering in the framework
of a left-right symmetric model (LRSM). Constraints on the LRSM parameters
$M_{Z_{2}}$ and $\phi$ were obtained. Based on new measurements
we present an update to these constraints and also include in the
calculation the radiative corrections.

\noindent pacs 13.10.+q, 12.15.Mm, 12.60.Cn, 14.60.Lm
\end{abstract}

\begin{center}
Submitted to {\bf Phys. Rev. D}
\end{center}

\newpage


In a previous paper \cite{1} we get constraints to the LRSM parameters
$M_{Z_{2}}$ and $\phi$ from the comparison between the theoretical prediction
of the weak coupling constants, $g_{{}_V}$ and $g_{{}_A}$, and the
experimental values reported at that time \cite{2}. Very recently these
quantities have been reported experimentally by the CHARM II
Collaboration \cite{3},

\bq
g^{\rm exp}_{{}_V} = -0.035 \pm 0.017
\eq
\bq
g^{\rm exp}_{{}_A} = -0.503 \pm 0.017,
\eq

\noindent these values for the coupling constants were obtained using
neutrino-electron and antineutrino-electron scattering experiments. We make
the comparison with our effective coupling constants
$\l(g^{e\nu}_{{}_V}\r)_{\rm LRSM}$ and
$\l(g^{e\nu}_{{}_A}\r)_{\rm LRSM}$ in a straightforward manner. However this
time we shall take into account the radiative corrections to make this
comparison as precise as possible.

The Hamiltonian for neutral currents at tree level, in the framework of the
LRSM, is given by

\bq
{\cal H}^{\rm eff}_{NC} = \frac{4G_F}{\sqrt2} \l[aJ^{z^+}_LJ^z_L
+b\l(J^{z^+}_L J^z_R +J^{z^+}_RJ^z_L\r) +cJ^{z^+}_RJ^z_R \r],
\eq

\noindent The radiative correction to this kind of models have been discussed
widely in the literature \cite{4}. The main conclusion that we can get from
this analysis is that taking Standard Model (SM) radiative corrections is
enough at this level of precision. This would make the previous Hamiltonian to
become

\bq
{\cal H}^{\rm eff}_{NC} = \frac{4G_F}{\sqrt2}
\l[a \bar J^{z^+}_L \bar J^z_L+b\l(\bar J^{z^+}_L J^z_R
+J^{z^+}_R\bar J^z_L\r) +cJ^{z^+}_RJ^z_R \r],
\eq

\noindent where the difference is that now
$\bar J^z_L = \frac12\bar\psi\gamma^{\mu}\l(\bar g_{{}_V}
-\bar g_{{}_A}\gamma^5\r)\psi$ and  $\bar g_{{}_V}$ and $\bar g_{{}_A}$
are the coupling constants including radiative corrections:

\begin{eqnarray}
\bar g_{{}_V} &=& \rho_{\nu e} \l(-\frac12
+ 2\kappa_{\nu e} sin^2 \theta_W \r) \\
\bar g_{{}_A} &=& \rho_{\nu e} \l(-\frac12 \r).
\end{eqnarray}

The values for $\rho_{\nu e}$ and $\kappa_{\nu e}$ are those given by the
Particle Data Group (PDG) \cite{5}, $\rho_{\nu e}=1.0131$ and
$\kappa_{\nu e}=1.0332$. This difference will make that our formulas change in
the following way

\bq
\l(g^{e\nu}_{{}_A}\r)_{\rm LRSM} = \frac12 G(1-\beta),
\eq
\[
\l(g^{e\nu}_{{}_V}\r)_{\rm LRSM} = \frac12 F(1-\alpha)
\]
\noindent where
\begin{eqnarray}
F &= &\frac12 \l(a\bar g_{{}_V} +b(\bar g_{{}_V} +g_{{}_V})+cg_{{}_V} \r) \\
G &= &\frac12 \l(a\bar g_{{}_A} +b(\bar g_{{}_A} -g_{{}_A})-cg_{{}_A} \r) \\
\alpha &= &\frac{-1}{2F} \l(a\bar g_{{}_V} -b(\bar g_{{}_V} -g_{{}_V})
-cg_{{}_V} \r) \\
\beta &= &\frac{-1}{2G} \l(a\bar g_{{}_A} -b(\bar g_{{}_A} +g_{{}_A})
+cg_{{}_A} \r)
\end{eqnarray}

We use the new experimental results to make a $\chi^2$ fit. We take the value
$sin^2 \theta_W = 0.2247$ reported by the PDG. The plot in Figure (1) shows
$\chi^2$ as a function of $M_{Z_2}$ and $\phi$ for values of $\chi^2$ below
0.18, which traduce to a 91.4 \% in the confidence level. The plot in
Figure (2) shows the allowed region for the LRSM parameters $M_{Z_2}$ and
$\phi$. Comparing to what we  obtained in the previous paper, we see that the
allowed region is larger in this case.

The results obtained are in agreement with the analysis done by the CHARM II
Collaboration \cite{3}, once we take into account the differences on the
parametrization ($\phi = - \theta_{LR}$ \cite{6}.)

We finally comment on the constraints to the neutrino mass obtained in the
previous paper. With the new allowed region for the LRSM parameters, we see
that the general behavior of the neutrino mass bound is the same and the upper
limit we get including radiative corrections is $m_{\nu} \leq 8.8$ KeV.

In conclusion we can say that radiative corrections make neglegible
contributions to this process and that the new experimental results allow the
LRSM parameters to have a wider range of variation.

This work was supported in part by CONACyT (M\'exico).


\vspace{2cm}


\noindent{\Large{\bf Figure Captions}}
\vspace{1cm}

\noindent Fig 1 Plot of $\chi^2$ for values below the cutoff 0.18 (91.4 \%
C. L.).

\vspace{.5cm}

\noindent Fig 2 Allowed region for $\phi$ and $M_{Z_2}$ at 91.4 \% C. L.

\vspace{.5cm}

\noindent Fig 3 Plot of the maximum value of the neutrino mass (muon or tau
neutrino) as a function of the LRSM parameters $\phi$ and $M_{Z_2}$.

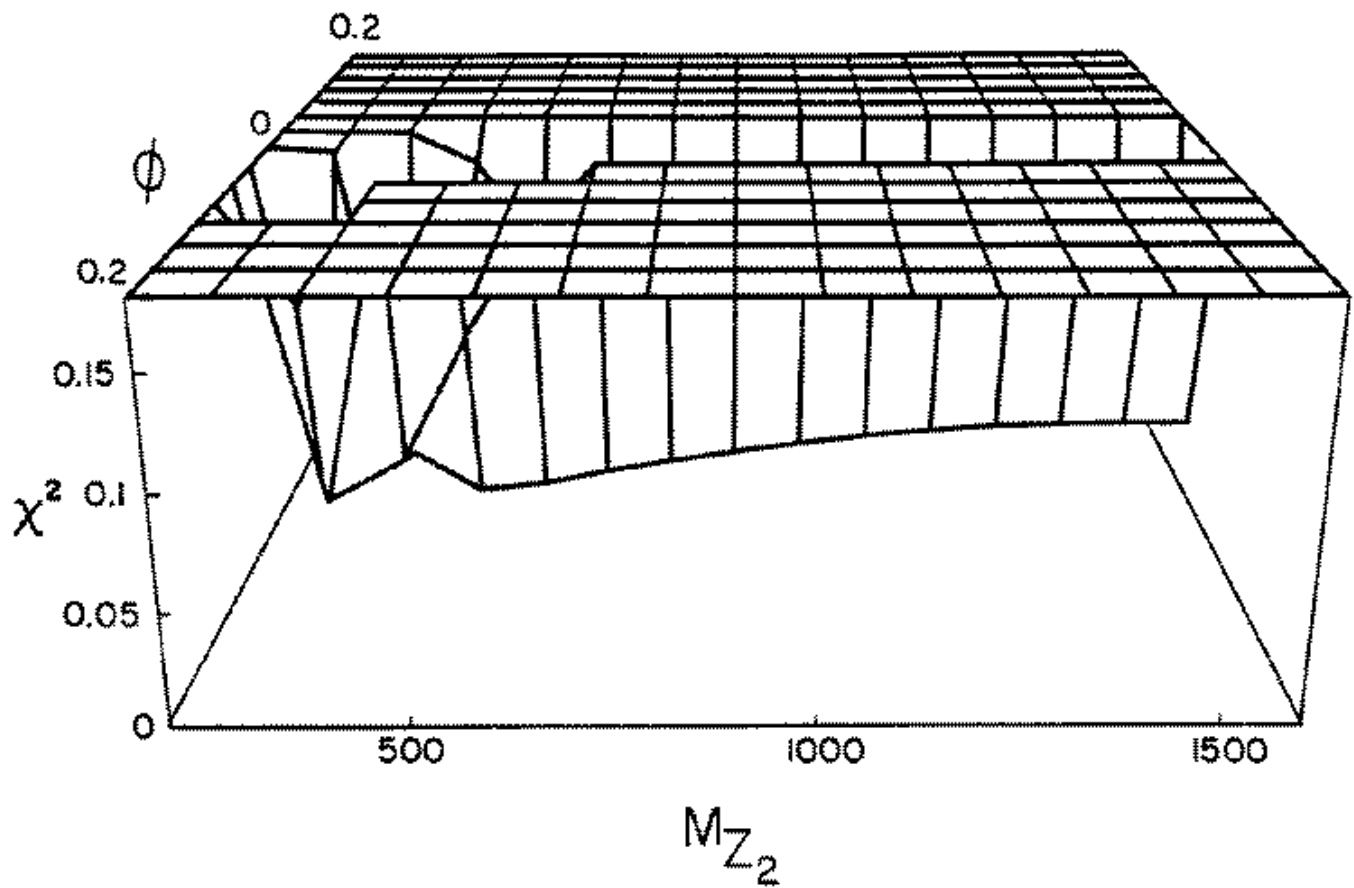
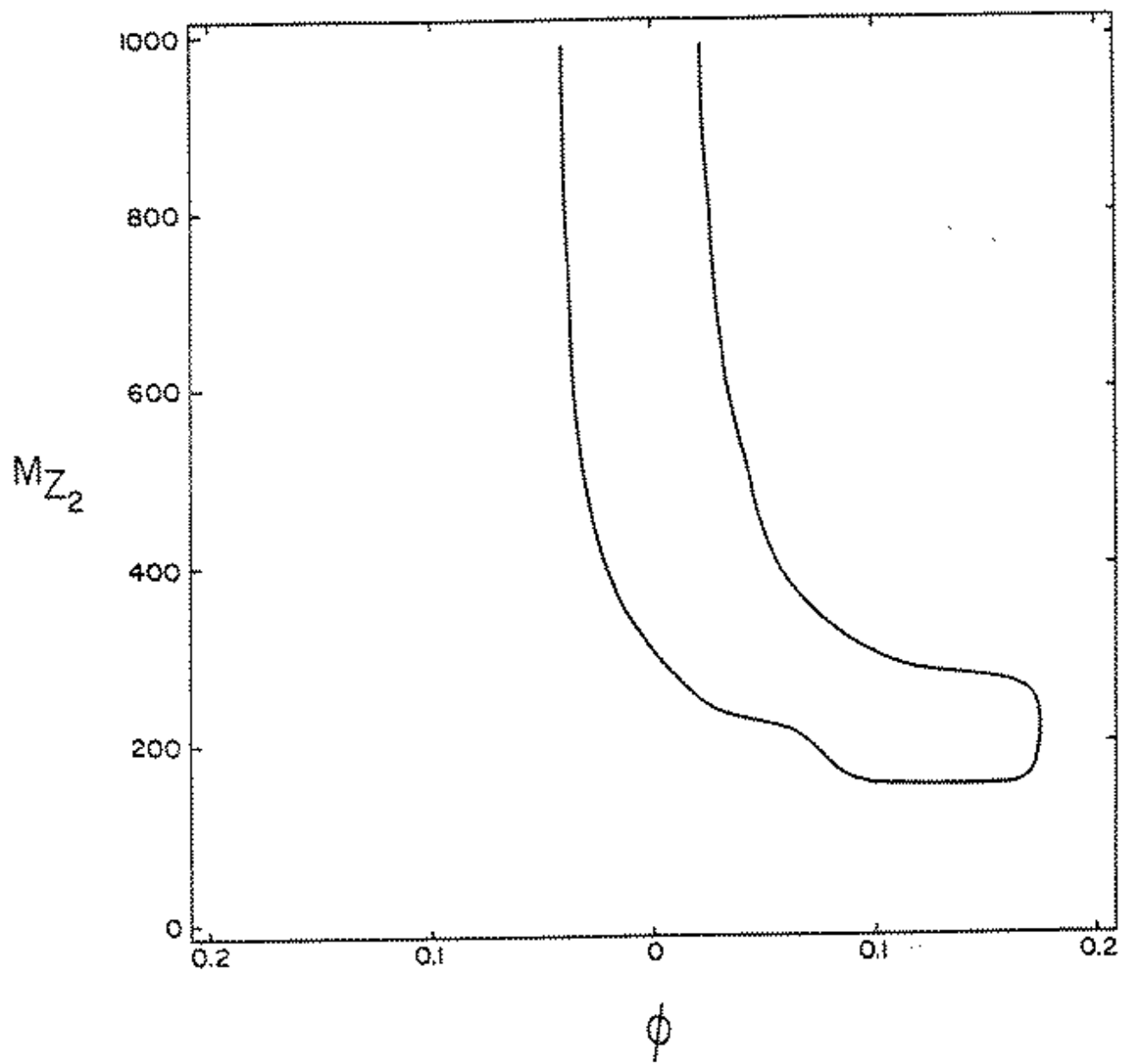
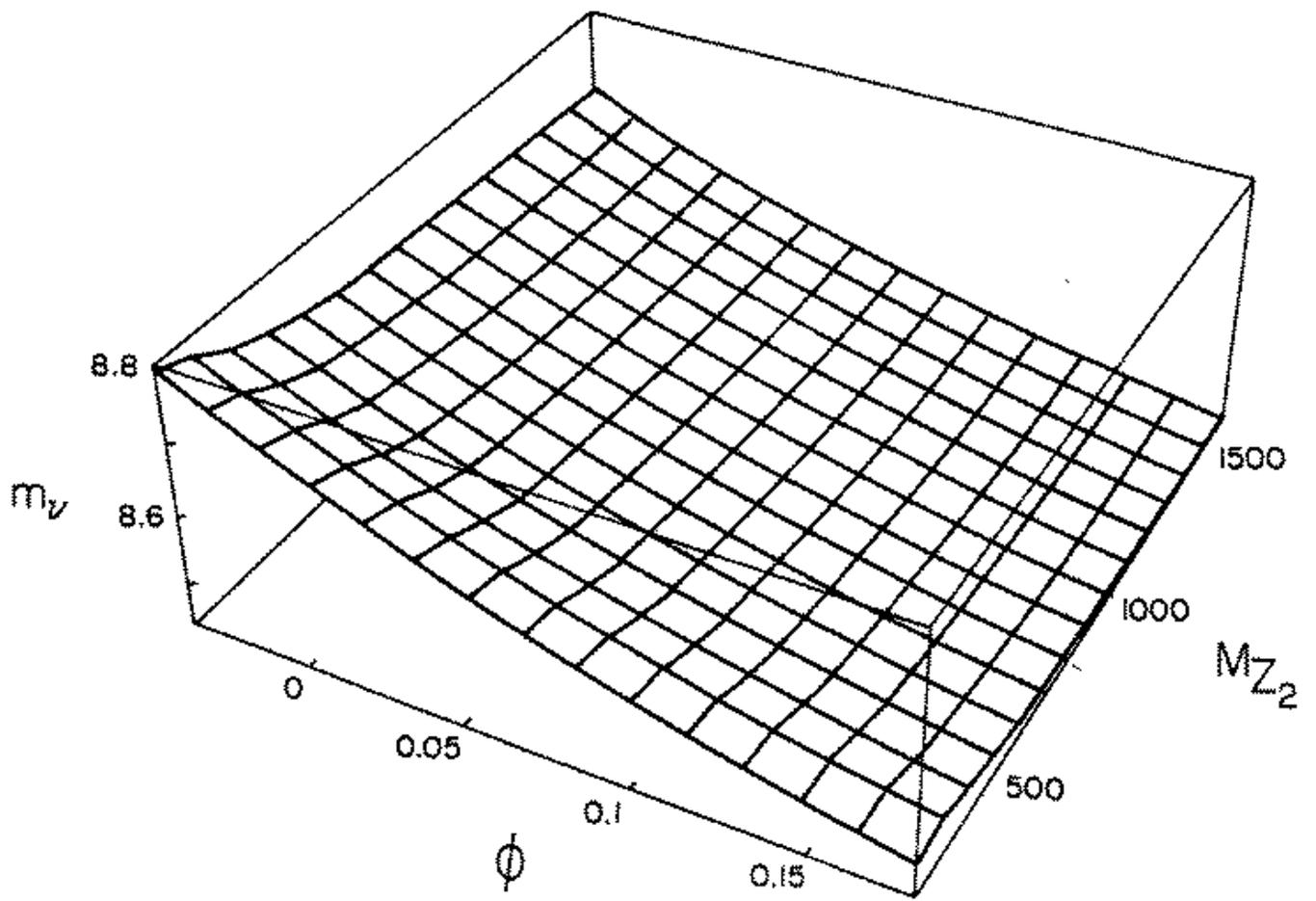
\end{document}